\documentclass[superscriptaddress,preprintnumbers,amsmath,amssymb,twocolumn,floats]{revtex4-1}
\usepackage{txfonts}
\usepackage{amssymb}
\usepackage{graphicx}
\usepackage{sidecap}
\usepackage{CJK}
\usepackage{txfonts}
\usepackage{url}
\usepackage{color}
\usepackage{amsmath}
\usepackage{soul}
\soulregister\cite7
\numberwithin{equation}{section}
\begin{document}

\title{Electronic structure of the high-mobility two-dimensional antiferromagnetic metal GdTe$_3$}

\author{J. S. Liu}
\thanks{Equal contributions}
\affiliation{Center for Excellence in Superconducting Electronics, State Key Laboratory of Functional Materials for Informatics, Shanghai Institute of Microsystem and Information Technology, Chinese Academy of Sciences, Shanghai 200050, China}
\affiliation{Center of Materials Science and Optoelectronics Engineering, University of Chinese Academy of Sciences, Beijing 100049, China}

\author{S. C. Huan}
\thanks{Equal contributions}
\affiliation{School of Physical Science and Technology, ShanghaiTech University, Shanghai 201210, China}

\author{Z. H. Liu}
\affiliation{Center for Excellence in Superconducting Electronics, State Key Laboratory of Functional Materials for Informatics, Shanghai Institute of Microsystem and Information Technology, Chinese Academy of Sciences, Shanghai 200050, China}
\affiliation{Center of Materials Science and Optoelectronics Engineering, University of Chinese Academy of Sciences, Beijing 100049, China}

\author{W. L. Liu}
\affiliation{Center for Excellence in Superconducting Electronics, State Key Laboratory of Functional Materials for Informatics, Shanghai Institute of Microsystem and Information Technology, Chinese Academy of Sciences, Shanghai 200050, China}

\author{Z. T. Liu}
\affiliation{Center for Excellence in Superconducting Electronics, State Key Laboratory of Functional Materials for Informatics, Shanghai Institute of Microsystem and Information Technology, Chinese Academy of Sciences, Shanghai 200050, China}

\author{X. L. Lu}
\affiliation{Center for Excellence in Superconducting Electronics, State Key Laboratory of Functional Materials for Informatics, Shanghai Institute of Microsystem and Information Technology, Chinese Academy of Sciences, Shanghai 200050, China}

\author{Z. Huang}
\affiliation{Center for Excellence in Superconducting Electronics, State Key Laboratory of Functional Materials for Informatics, Shanghai Institute of Microsystem and Information Technology, Chinese Academy of Sciences, Shanghai 200050, China}

\author{Z. C. Jiang}
\affiliation{Center for Excellence in Superconducting Electronics, State Key Laboratory of Functional Materials for Informatics, Shanghai Institute of Microsystem and Information Technology, Chinese Academy of Sciences, Shanghai 200050, China}

\author{X. Wang}
\affiliation{School of Physical Science and Technology, ShanghaiTech University, Shanghai 201210, China}
\affiliation{Analytical Instrumentation Center, School of Physical Science and Technology, ShanghaiTech University, Shanghai 201210, China}

\author{N. Yu}
\affiliation{School of Physical Science and Technology, ShanghaiTech University, Shanghai 201210, China}
\affiliation{Analytical Instrumentation Center, School of Physical Science and Technology, ShanghaiTech University, Shanghai 201210, China}

\author{Z. Q. Zou}
\affiliation{School of Physical Science and Technology, ShanghaiTech University, Shanghai 201210, China}
\affiliation{Analytical Instrumentation Center, School of Physical Science and Technology, ShanghaiTech University, Shanghai 201210, China}

\author{Y. F. Guo}
\email{guoyf@shanghaitech.edu.cn}
\affiliation{School of Physical Science and Technology, ShanghaiTech University, Shanghai 201210, China}

\author{D. W. Shen}
\email{dwshen@mail.sim.ac.cn}
\affiliation{Center for Excellence in Superconducting Electronics, State Key Laboratory of Functional Materials for Informatics, Shanghai Institute of Microsystem and Information Technology, Chinese Academy of Sciences, Shanghai 200050, China}
\affiliation{Center of Materials Science and Optoelectronics Engineering, University of Chinese Academy of Sciences, Beijing 100049, China}


\date{\today}

\begin{abstract}

The new found two-dimensional antiferromagnetic GdTe$_3$ is attractive owing to its highest carrier mobility among all known layered magnetic
materials, as well as its potential application for novel magnetic twistronic and spintronic devices. Here, we have used high-resolution angle-resolved photoemission spectroscopy to investigate its Fermi surface topology and low-lying electronic band structure. The Fermi surface is partially gapped by charge density wave below the transition temperature, the residual part reconstructs making GdTe$_3$ metallic. The high carrier mobility can be attributed to the sharp and nearly linear band dispersions near the Fermi energy. We find that the scattering rate of the linear band near the Fermi energy is almost linear within a wide energy range, indicating that GdTe$_3$ is a non-Fermi liquid metal. Our results in this paper provide a fundamental understanding of this layered Van der Waals antiferromagnetic materials to guide future studies on it. 

\end{abstract}

\maketitle


\section{\label{sec:level1}Introduction}
Two-dimensional (2D) van der Waals (vdW) magnets have always been pursued not only they can provide extraordinary opportunity to investigate the magnetism in 2D limit, but also their easy exfoliation into multi-/mono-layer that would facilitate applications for developing various novel devices, such as atomically thin magneto-optical devices, energy-efficient magnetoelectric devices and on-chip optical communication and quantum computing~\cite{2Dreview1,2Dreview2,2Dreview3}. In principle, strong thermal fluctuation in 2D magnets with isotropic exchange would suppress the intrinsic long-range magnetic order by prohibiting spontaneous symmetry breaking. However, when magnetic anisotropy exists, the long-range magnetic order can still survive even in exfoliated mono-layer 2D vdW magnets, such as CrI$_3$ and Cr$_2$Ge$_2$Te$_6$ reported recently~\cite{2D2,2D3}. Since these discoveries, continuous research interest has been inspired in this intriguing family of materials ~\cite{2D1,2D2,2D3,2D4}. So far, most discovered 2D vdW magnets are limited to a few insulating or semiconducting materials, while conducting ones are still rare, especially for those with the high carrier mobility, which are actually critical for building magnetic twistronic and spintronic devices~\cite{Cao2018,2Dreview3}. 

Recently, GdTe$_3$ has been reported as a potential conducting 2D antiferromagnetic (AFM) vdW material with the highest carrier mobility among all known layered magnetic materials~\cite{GDtransport}. It can be mechanically exfoliated into flakes with only several atomic layers and then preserves the long-range AFM order. Actually, as a rare-earth tritelluride, GdTe$_3$ bears some other exotic electronic properties, such as the incommensurate charge-density-wave (CDW) ground state and superconductivity under high pressure~\cite{GDtheory,SC,SC1,pressure,SC2}. To understand the underlying mechanism of these unique properties, the low-energy electronic structure is crucial~\cite{2DARPES,2DARPES1}, which is as well of important directive significance to further device development based on 2D magnets. Note that soft x-ray angle-resolved photoemission spectroscopy (ARPES) measurement has been performed in Ref [9]. However, this key information of GdTe$_3$ is still absent to date.

In this work,  high-resolution ARPES and magnetic transport measurements were performed to achieve in-depth insights into the superior physical properties of GdTe$_3$. The CDW wave vector, determined from the gapped region of Fermi surface (FS), is estimated to be about 5/7$\times$2$\pi$/b. The residual Fermi surface reconstructs and diamond-like Electron pockets form at the Brillouin zone (BZ) boundary by the crossing of main bands and its folded counterparts, which contribute to metallic behavior even below the CDW transition temperature. The high mobility found in GdTe$_3$ could be attributed to the steep linearly dispersive bands crossing $E_F$ at the ungapped FS region, which extrapolate a rather small effective mass of the carriers. Through a close inspection of the line-width dependence of the linear band, non-Fermi liquid properties are also revealed in GdTe$_3$. lastly, the antiferromagnetic order in GdTe$_3$ was also discussed.

\section{Experiments}

\begin{figure*}
   \centering
   \includegraphics[scale=1]{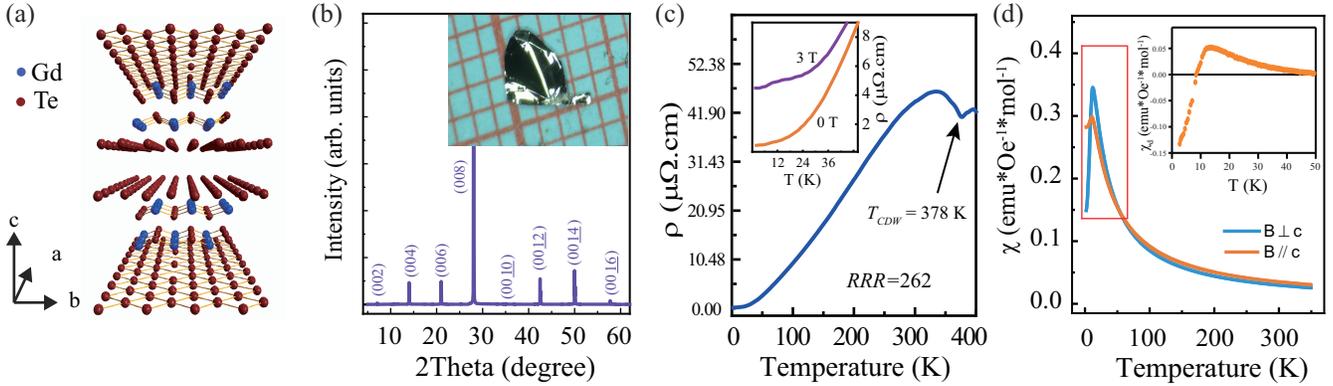}
   \caption{\label{figure:wide}(a) The crystal structure of GdTe$_3$. (b) Powder X-ray diffraction pattern of cleaved GdTe$_3$ single crystals, inset of (b) shows the optical image of a typical single crystal sample at the millimeter scale. (c) Temperature dependence of the in-plane resistivity of the sample. The inset is the in-plane resistivity under transverse magnetic fields of 0 and 3 T at low temperature range. (d) Temperature-dependent magnetic susceptibility of a bulk GdTe$_3$ crystal,inset is zoom in of the red box.}
\end{figure*}

GdTe$_3$ single crystals were synthesized via the self-flux method. The starting materials Gd and Te were mixed in a molar ratio of Gd : Te = 1 : 25 and then placed into an alumina crucible which was then sealed in the quartz tube. The assembly was heated up to 1273 K in a furnace, maintained for 15 hours and was subsequently cooled down to 823 K at a temperature decreasing rate of 3 K/hour. The Te flux was removed by quickly placing the tube into a high speed centrifuge at this temperature and GdTe$_3$ single crystals were left in the alumina crucible. High-resolution ARPES measurements were carried out at 03U beamline of Shanghai synchrotron radiation facility (SSRF). The ARPES system is equipped with a Scienta Omicron DA30 electron energy analyzer. The energy and angular resolutions were set to around 10 meV and 0.2$^\circ$, respectively. Clean surfaces for ARPES measurements were obtained by cleaving samples in ultrahigh vacuum better than 5.0 $\times$ 10$^{-11}$ Torr. During measurement, the temperature was kept at 15 K. The temperature dependent in-plane resistivity $\rho$~(T) was measured in a four-probe configuration under various magnetic fields up to 3 T in a commercial physical property measurement system from Quantum Design (QD). The direct current (dc) magnetic susceptibility along the in-plane and out-of-plane directions with a magnetic field of 1 kOe in the temperature range of 1.8 to 350 K were measured in the magnetic property measurement system from QD.  

\section{Results and Discussion}

Figure 1(a) shows the crystal structure of GdTe$_3$, which is of a quasi-2D layered orthorhombic structure with the space group $Cmcm$ (No. 63). It consists of double puckered Gd-Te slabs that are sandwiched by two Te square-net sheets. Thus, it can be mechanically exfoliated into flakes due to the weak vdW binding between two neighboring Te sheets, and its cleaved surface is parallel to the $ab$-plane. Note that we define the out-of-plane direction of GdTe$_3$ along the $c$-axis, which is different from some previous reports on other RTe$_3$ materials where the out-of-plane direction is defined as $b$-axis~\cite{CeTe3,ErTe3}. Fig.1(b) shows the x-ray diffraction (XRD) pattern of the as-grown GdTe$_3$ crystal. It exhibits well defined sharp diffraction peaks that can be indexed by the (00l) reflections, indicating single crystalline in nature. The inset of Fig.1(b) is the optical image of GdTe$_3$ single crystalline sample in a millimeter scale, which shows a flat and shiny surface after cleave.

The typical temperature dependence of the in-plane resistivity measured in various transversely applied magnetic fields ranging from 0 to 3T are shown in Fig. 1(c). In zero magnetic field, the resistivity shows metallic behavior with $\rho$~(300K) = 4.454 $\mu\Omega\cdot cm$ and $\rho$~(2.5K) = 0.017 $\mu\Omega\cdot cm$. The residual resistance ratio (RRR), which is defined as $\rho_{xx}$(300 K)/$\rho_{xx}$(2.5 K), can reach as high as 262, indicating the high quality of our samples with less crystallographic defects and impurities. The up-turn of low-temperature resistivity under the applied field of 3 T signifies the AFM order, as demonstrated in the inset of Fig.1(c). The AFM order was confirmed by the temperature-dependent magnetic susceptibility of bulk GdTe$_3$, as shown in Fig.1(d). Our result illustrates that the N$\rm\acute{e}$el transition temperature ($T_{N}$) is around 13 K. We find that $\chi$(T) is almost isotropic in the paramagnetic state regardless of the direction of applied magnetic field, which is consistent with the S-state (L=0) of the Gd$^{3+}$. In contrast, below $T_N$, $\chi_{ab}$(T) drops much more evidently with the decrease of temperature than $\chi_{c}$(T), suggesting that the Gd moment orders mainly in the basal $ab$-plane.Note that the AFM in GdTe3 are mainly from the Gd moment ordering, which is different with the magnetism is promoted by the presence of defects in semiconducting molybdenum based Van der Waals systems 2H-MoTe$_2$ and MoSe$_2$~\cite{magdefect}. The temperature dependence of the magnetic susceptibility shows the value of $\chi_{ab}$ and $\chi_{c}$ are substantially different in the magnetic transition temperature range, which is deviate from the classic picture of an ideal antiferromagnet~\cite{cantedAFM}. This behavior can be explained by invoking a transition from the high temperature paramagnetic state to a canted-AFM state and to a low temperature AFM state, similar in layered Fe-based system EuFe$_2$As$_2$~\cite{cantedAFM}. Here, such the magnetic anisotropy might result in the existing 2D magnetism in GdTe$_3$ even after being exfoliated into flakes with only several atomic layers~\cite{Gibertini2019Review}.

\begin{figure*}[!htpb] 
    \centering
    \includegraphics[scale=1]{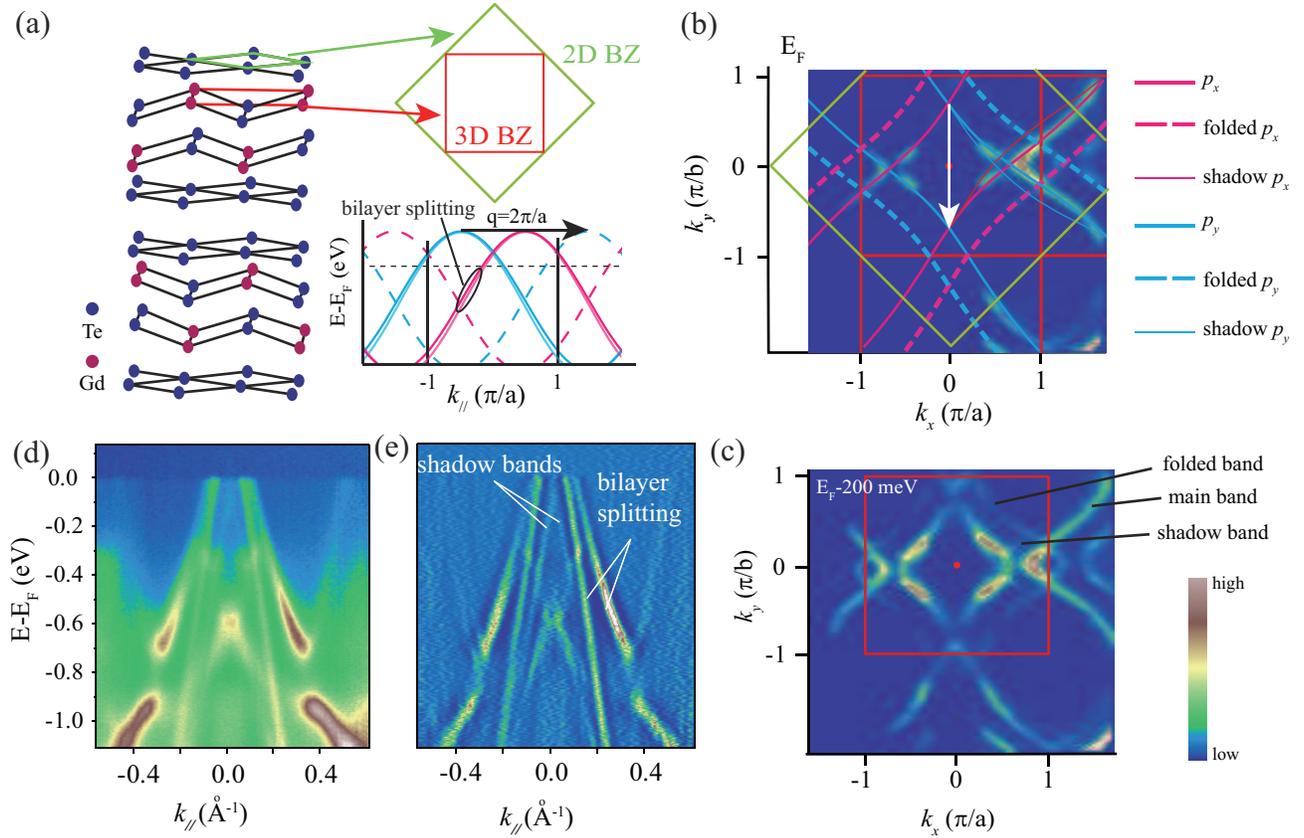}
    \caption{(a) The sketch of 3D BZ and 2D BZ, which result from 3D unit cell and Te square-net sheet, respectively.The lower right panel of (a) is the schematic of the dispersion for folded band and bilayer splitting. (b) ARPES FS map obtained by integration of spectral weight in a 10 meV window below $E_F$, taken with photon energy of 95 eV and at 15K. The arrow indicate the wave vector of CDW. (c) Photoemission integrated intensity plot at binding energy of 0.2 eV within a 10 meV window. (d) Band dispersion near $E_F$ measured with the photon energy of 33 eV. (e) The corresponding second derivative image of (d) with respect to the momentum.}
    \label{fig:my_label}
\end{figure*}

Next, Fermi surface and the low-lying band structures were explored by high-resolution ARPES. According to the crystal structure of GdTe$_3$, two different Brillouin zones (named 3D BZ and 2D BZ, respectively) are obtained in the reciprocal space, as sketched in Fig.2(a). The 3D BZ is produced by the stacking of puckered GdTe slabs, and the 2D BZ is built based on Te square-net sheets. The 3D BZ is reduced by half and rotated by 45$^{\circ}$ from the 2D one. As shown in Fig.2(b), we present the Fermi surface map of GdTe$_3$ obtained by integration of spectral weight within a 10 meV window around $E_F$. Similar to previous literature data on other RTe$_3$ materials, in-plane $p_x$ and $p_y$ orbitals of Te atoms from square nets dominate electronic states near $E_F$~\cite{CeTe3,ErTe3,CPL1,PrTe3}. Three electrons are donated per Gd ion in GdTe$_3$, two of these denoted electrons fill the Te 5$p$ bands of the GdTe slabs, while one electron transfer to the Te square-net sheets. This result in the GdTe slabs are insulating and the Te square-net sheets are conductive for these 5$p$ bands which are halffilled. It is plausible to achieve a good description of the electronic structure with the 2D tight-binding calculation based on the Te square-net plane. However, there still exist some extra Fermi surface portions which could not be accounted by the calculation. In fact, to acquire the 3D lattice symmetry and satisfy the equivalency of the first and second 3D BZs, additional folded bands which arise from the bands in 2D BZ folded back into the 3D BZ are also observed in the reduced 3D BZ. As a whole 3D crystal, the charge density from the underlying GdTe layers indeed impinges upon Te square-net planes, serving as a commensurate potential that scatters the electrons therein, as sketched in the lower right panel of Fig. 2(a), suggesting the sizable coupling between Te square-net and the Gd-Te magnetic slab. There is no spectral weight in the vicinity of \textit{k$_x$} = 0 axis at $E_F$, indicating the presence of a CDW gap in this region. The wave vector of CDW determined from the FS is close to 5/7 $k_y$, demonstrated by the arrow line in Fig.2(b), in a good agreement with the value obtained by the scanning tunneling microscope measurement~\cite{GDtransport}. 
Fig. 2(c) displays a map of the spectral weight integrated in a 10 meV window at the binding energy of 200 meV. The zero intensity of spectral weight at \textit{k$_x$} = 0 suggests the maximum of CDW gap is larger than 200 meV in GdTe$_3$, in agreement with that the CDW transition temperature is well above room temperature. The CDW shadow bands can be clearly observed at high \textit{k$_x$} region, as indicated in Fig. 2(d) and its corresponding second derivative image (e). Note that the two clear narrow lines reaching for $E_F$ result from bilayer splitting as there are two Te planes per unit cell.

\begin{figure}
    \centering
    \includegraphics[width=8.5cm]{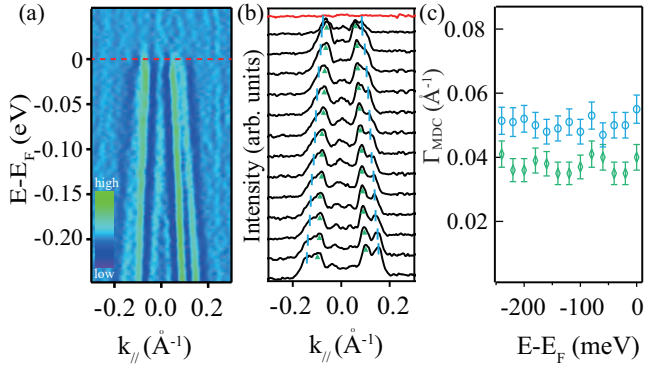}
    \caption{Band dispersion near $E_F$ taken along $k_y$ at $k_x$=0.26 $\AA^{-1}$ and with the photon energy of 33 eV. (a) is second derivative image and its corresponding MDCs (b). A linear fit was used to evaluate its Fermi velocity. The red MDC line is recorded at $E_F$. (c) Extracted full width of half maximum of the main band in momentum by using Lorentz-fitting.}
    \label{fig:my_label}
\end{figure}

Lei $et~al.$ reported that GdTe$_3$ bears the highest carrier mobility among all known layered magnetic materials, which is even comparable to some topological semimetals \cite{ZrSiSNC,PANS}. To address the extremely large carrier mobility in GdTe$_3$, its detailed band dispersion traversing $E_F$ was further investigated. Fig. 3(a) shows the second derivative photoemission plot with respect to energy down to the binding energy over 0.2 eV taken along $k_x$ at $k_y$ = 0.26 $\AA^{-1}$ with the photon energy of 33 eV. It shows rather steep and linear dispersions of bilayer-splitted conduction main bands persisting in a wide energy window, revealing the rather low effective masses of these bands. The Fermi velocity (the slope of the band dispersion) could be accurately extracted by a linear fit to the corresponding momentum distribution curves (MDCs) [Fig. 3(b)]. Two 
Fermi velocities were extracted from the bilayer-splitted main bands, which are 0.68 $\times$ 10$^6$ and 1.03 $\times$ 10$^6$ m/s, respectively. For the dispersioin which is linear, we can further deduce the effective mass of bands to be nearly zero. Since the mobility is inversely proportional to both the scattering rate and effective mass of carriers in conducting materials, the  low effective mass of carriers in GdTe$_3$ would be conductive to improve its mobility. 
As for the scattering rate in GdTe$_3$, it is proportional to the imaginary part of the complex self-energy $\Sigma$(\textbf{k},$\omega$), which can be readily obtained from the Lorentzian linewidth of the ARPES MDC in the k-independent approximation~\cite{ARPESselfenergy1,ARPESselfenergy2,NFL3,NFL4}. In this quasiparticle scheme, there are three main processes that dominate the scatting rate (linewidth of the MDC): decay by the electron–hole (e–h) pair generation, electron–phonon (e–ph) coupling and emission of collective charge excitations via electron–plasmon (e–pl) coupling~\cite{ARPESselfenergy2}. For conventional 3D Fermi liquid systems, according to the Landau's theory, the scattering rate from the e–h pair generation increases as $\omega^{2}$ away from $\omega$=0 ($E_F$); while the scattering process occurs in the pure two dimension, the quadratic energy dependence would be modified by a logarithmic factor~\cite{FL2D}. Fig. 3(c) shows the full width at half maximum (FWHM) of MDCs of GdTe$_3$ as the function of binding energy. Here, these FWHMs of MDCs are obtained by using a constant flat background with Lorentzian fitting. Surprisingly, both MDCs' linewidths show well linear relationship with the binding energy within a wide energy range, which apparently deviates from either the 3D or 2D Fermi liquid behavior. In fact, there are many compounds showing non-Fermi liquid behaviors mainly due to correlation effects, such as superconductors\cite{NFL5,NFL6,NFL7}, Dirac semimetals\cite{NFL3,NFL8}, layered magnets\cite{NFL4}, but the underlying physics need to be further studied. Such unusual behavior has also been reported in some typical 2D materials such as graphene and layered electron gas~\cite{nonFL,plasmaonin2D}. In graphene, the departure of the self-energy from ordinary metallic system was explained by the electron-electron interaction and the diminishing e–ph interaction when the semimetallic regime is approached~\cite{nonFL}. As for layered electron gases, such a behavior has been attributed to the significant e-pl coupling ~\cite{plasmaonin2D1,plasmaonin2D2}. In 2D materials mentioned above, electrons can move freely in the confined layers, while the interlayered tunneling is forbidden. Under this condition, plasmon bands are formed for plasmons associated with the individual layers couple via long-range Coulomb interactions. Thus, the e-pl interaction becomes an effective channel for excited electron relaxation even at small excitation energies. 
In GdTe$_3$,the GdTe slab is quite insulating and the conducting electrons are well confined in Te square-net sheets, which is of typical such layered electron gas model. Hence, we can reasonably conclude that the linear increase of the linewidth in GdTe$_3$ would be attributed to the electron-electron interaction and diminishing e–ph interaction like in graphene or significant e-pl interaction like in layered electron gas systems. 

\begin{figure}[!htpb]
     \centering
     \includegraphics[scale=1]{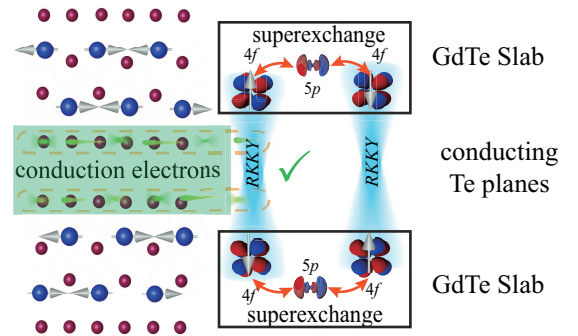}
    \caption{Schematic of rare-earth antiferromagnetic superexchange-like and RKKY interaction.}
    \label{fig:my_label}
\end{figure}

Finally, we would like to discuss the long-ranged antiferromagnetic order in GdTe$_3$. Some previous works have pinned down the relationship between the CDW formation and magnetic order in RTe$_3$\cite{magneticofRTe1,magneticofRTe2,magneticofRTe3,magneticofRTe4}. The ordered magnetic moment lies within the RTe slab, the dominant magnetic interaction among the R$^{3+}$ ions is direct exchange or superexchange interaction. However, Shubnikov–de Haas (SdH) oscillations measurement revealed that magnetic order in GdTe$_3$ has a significant interplay with its conducing electrons, because the amplitude of oscillations showed a clear deviation from the Lifshitz-Kosevich formula and reached a plateau in the magnetically ordered regime~\cite{GDtransport}. Other magnetic property studies as well suggested the conducting electron plays an role in long-ranged antiferromagnetic order, like the magnetic transition temperature in conductive GdTe$_3$ higher than that in semimetallic GdTe$_2$ \cite{magneticofRTe3,RTe2}. From the ARPES results, we can clearly observe the folding bands, although their intensity is relatively lower compared with the original bands. Based on well-accepted principle that the intensity of a folded FS in ARPES is proportional to the coupling responsible of the folding, the visible intensity of the folding bands in GdTe$_3$ means the coupling between Te square-net and the GdTe magnetic slab is not much weak. Then, we could conclude that the dominant magnetic interactions among the Gd$^{3+}$ ions are the magnetic dipole interaction or the superexchange interaction through chalcogen atoms in GdTe slab, while the Ruderman-kittel-kasuya-Yosida (RKKY) type interaction plays a second but not negligible role. 

\section{conclusion}
In conclusion, we have presented a detailed ARPES study on the high-mobility 2D conducting antiferromagnetic metal GdTe$_3$. Te 5$p$ bands derived from Te square nets dominate the low-lying electronic structure. The high mobility found in GdTe$_3$ is demonstrated to be closely related with the steep and nearly linear band dispersion traversing $E_F$. 
The antiferromagnetic ordering in GdTe$_3$ are dominated by the magnetic dipole interaction or the superexchange interaction, while RKKY-type interaction promotes the antiferromagnetic ordering. Our work in this paper help people understand the physical properties of this layered Van der Waals conductive antiferromagnetic materials from electronic structure perspective guiding future low-power consuming magnetic device design on it.

\begin{acknowledgments}
This work was supported by the National Key R$\&$D Program of the MOST of China (Grant No. 2016YFA0300204), the National Science Foundation of China (Grant Nos. 11404360, 11874264 and 11574337), and the Natural Science Foundation of Shanghai (Grant No. 14ZR1447600). Y. F. Guo acknowledges the starting grant of ShanghaiTech University and the Program for Professor of Special Appointment (Shanghai Eastern Scholar). Part of this research used Beamline 03U of the Shanghai Synchrotron Radiation Facility, which is supported by ME2 project under contract No. 11227902 from National Natural Science Foundation of China. D.W.S. is supported by ``Award for Outstanding Member in Youth Innovation Promotion Association CAS''. The authors also thank the support from the Analytical Instrumentation Center (SPST-AIC10112914), SPST, ShanghaiTech University.
\end{acknowledgments}

\nocite{*}

\bibliography{GdTe3}

\end{document}